\documentclass{ws-ijmpa}
\usepackage{cite}
\usepackage{amsfonts}
\usepackage{amsmath}
\usepackage{amssymb}

\newcommand{\be}{\begin{equation}}\newcommand{\ee}{\end{equation}}
\newcommand{\bea}{\begin{eqnarray}}\newcommand{\eea}{\end{eqnarray}}
\newcommand{\beaa}{\begin{eqnarray}}\newcommand{\eeaa}{\end{eqnarray}}
\newcommand{\ba}{\begin{array}}\newcommand{\ea}{\end{array}}
\newcommand{\bit}{\begin{itemize}}\newcommand{\eit}{\end{itemize}}
\newcommand{\ben}{\begin{enumerate}}\newcommand{\een}{\end{enumerate}}

\def\lab{\label}
\def\lan{\langle}

\def\ran{\rangle}
\def\rar{\rightarrow}

\begin{document}

\title{A novel variational approach for Quantum Field Theory:
example of study of the ground state and phase transition
in Nonlinear Sigma Model}

\author{Yuriy Mishchenko and Chueng-Ryong Ji}
\address{Physics Department \\
 North Carolina State University, Raleigh, NC }
\maketitle

\begin{abstract}
We discuss a novel form of the
variational approach in Quantum Field Theory 
in which the trial quantum
configuration is represented directly in terms
of relevant expectation values rather than, e.g.,
increasingly complicated
structure from Fock space.
The quantum algebra imposes constraints
on such expectation values so that
the variational problem is formulated here as an
optimization under constraints.
As an example of application
of such approach we consider the study
of ground state 
and critical properties in a variant of nonlinear sigma model.
\end{abstract}


The variational approach is one of the corner-stones of nonperturbative methods
in Quantum Mechanics and Quantum Field Theory. In this approach, the
expectation value of the Hamiltonian is analyzed on a set of quantum 
configurations of specific form and its minimum is sought.
Variational Method takes roots in Ritz Theorem \cite{ritz}
which states 
that for a hermitian Hamiltonian operator with the spectrum bounded from below 
\be\lab{E1}
\lan H \ran = \frac{\lan \Phi | H | \Phi \ran}{\lan \Phi | \Phi \ran}
\geq E_0,
\ee
for any quantum state $\Phi$.
Here $E_0$ is the lowest eigenvalue of $H$.
Ritz theorem can be transparently motivated using the fact that
hermitian operator should have a complete set of eigenstates,
i.e. an arbitrary quantum state can be represented as a linear
superposition of the Hamiltonian eigenstates. Then, after 
doing a simple algebra one can get
\be\lab{E2}
\lan \Phi | H | \Phi \ran
=\sum\limits_{n,n'} \phi_n^* \phi_{n'} \lan \psi_n | H |\psi_{n'} \ran 
=\sum\limits_{n} |\phi_n|^2 E_n \lan \psi_n | \psi_n\ran \geq E_0 \lan \Phi |\Phi \ran.
\ee
Here $|\Phi \ran = \sum\limits_{n}\phi_n |\psi_n \ran$ and
$|\psi_n\ran$ are eigenstates of Hamiltonian $H$ with
eigenvalues $E_n\geq E_0$.
From this simple argument, it follows that
one will get
an upper limit for the ground state energy
if one computes the Hamiltonian expectation value on  
arbitrary quantum state $\Phi$.
This suggests a way to generate a useful estimate
for the ground state energy by
considering a class of trial quantum states
$\{ \Phi_{\alpha} , \alpha\in {\bf A} \}$ and 
minimizing
$E(\alpha) = 
\frac{\lan \Phi_\alpha | H | \Phi_\alpha \ran}
{\lan \Phi_\alpha | \Phi_\alpha \ran}$
with respect to parameter $\alpha$.

The variational approach found a wide range of
applications due to its simplicity, possibility
of analytical results and important nonperturbative content.
One of the simplest examples is the problem of 
Hydrogen-like atom which is described in dimensionless
unitsby the Hamiltonian
\be\lab{E3}
H = -\frac 12 \nabla^2 - \frac 1 r.
\ee
Choosing an  exponential trial wavefunction
$\Phi_\alpha(r) = e^{-\alpha r}$,
one trivially obtains
\be\lab{E4}
\lan H \ran = \frac 12 \alpha(\alpha-2)
\ee
from which the solution $\alpha = 1$, $E=-0.5$ immediately
follows. In this case, the variational estimate is in fact the exact answer.

Many examples of the original applications of variational method
in Quantum Mechanics for computation of the energy levels
of simple molecules can be found in the literature.
In the later time, variational method had been employed  increasingly
for the precision many-body
calculations of  the ground state energies and first excited levels
of atoms and simple molecules. In theoretical studies, 
variational method is often used to derive 
important nonperturbative constraints,
e.g. asymptotic properties, on quantum-mechanical models.
First work using variational method in Quantum Field Theory dates back to
1960th since it had been applied to a
variety of atomic, nuclear and quark problems.
In QFT, variational method is commonly 
used to investigate stability
of the vacuum as well as to study dynamics of 
one- or two-particle excitations \cite{cole}.
Also, the conclusion about possibility 
of color-superconductivity in QCD vacuum can be justified
by considering the trial states with
quark-quark vacuum condensations.
The use of variational principle to formulate
nonperturbative equations of state for few-body systems
had also enjoyed growing popularity \cite{dare,dare01}.

In QFT, the variational principle is derived from the equation for
the covariant mass
\be\lab{E5}
P^\mu P_\mu |\Phi \ran = M^2 |\Phi \ran,
\ee
where $P^\mu$ is the momentum-energy operator.
In the Center of Momentum Frame, $\vec P |\Phi \ran =0$, 
this becomes
\be\lab{E6}
\ba{l}
(P^0)^2 |\Phi \ran = M^2 |\Phi \ran, \\
P^0 |\Phi \ran = \pm M |\Phi \ran.
\ea
\ee
Eq.(\ref{E6}) describes two energy spectra identical up to the
change of sign which are conventionally associated with
particle and antiparticle sectors of a field-theoretical model.
By excluding spectrum of negative energies, one arrives at
\be\lab{E7}
\frac {\lan \Phi | P^0 | \Phi \ran}
{\lan \Phi |  \Phi\ran} = E[\Phi] \geq E_0
\ee
or, equivalently, 
\be\lab{E8}
\lan \delta\Phi| H - E | \Phi \ran = 0.
\ee
Variational principle (\ref{E8}) applied to the states
from many-body Fock space
\be\lab{E9}
|\Phi \ran = \sum\limits_n 
\int\!\!\!\int\!\!\!\int 
dk_1\dots dk_n C_{(i)}(k_1,\dots,k_n) a_{k_1}^\dag \dots a_{k_n}^\dag
|0\ran
\ee
typically leads to an infinite system of coupled equations for the amplitudes
$C_{(i)}(k_1,\dots,k_n)$. This is usually simplified using
Tamm-Dancoff truncation \cite{tamm} or a variational ansatz \cite{dare}. Ansatzes 
involving a handful of few-body excitations, e.g.
$$
|\Phi\ran = ( 1 + \int dk f_k a^\dag_k + 
\int\!\!\! \int dk_1 dk_2 f_{k_1k_2} a^\dag_{k_1}a^\dag_{k_2} ) |0\ran,
$$
are often used in addition to ansatzes defined by coherent states
\be\lab{E10}
|\Phi \ran = \exp (\int dk f_ka^\dag_k)|0\ran
\ee 
or
\be\lab{E11}
|\Phi \ran = \exp (\int\!\!\! \int dk_1 dk_2 f_{k_1k_2}a^\dag_{k_1}a^\dag_{k_2})|0\ran.
\ee
While variational approach in QFT retains such important features
as simplicity in application and formulation as well as significant nonperturbative
contents, it suffers from numerous drawbacks which
hampered its applications in QFT. Among these
is the lack of explicit covariance, the necessity of nonperturbative
regularization/renormalization if the model is not a-priory
finite and the growing complexity of trial quantum states.
In this paper, we are going to consider a new variational approach
that may be helpful in dealing with the latter problem. 

We note that, while in Quantum Mechanics
the quantum state could be described by a single wavefunction,
in QFT one needs an infinite number of many-body
amplitudes $C_{(i)}(k_1,\dots,k_n)$, as introduced in Eq.(\ref{E9}),
to  properly
describe the quantum state.
This makes general variational problem intractable and even
simpler variational ansatzes, like in Eqs.(\ref{E10}) and (\ref{E11}),
require significant algebraic work to obtain  higher particle number
contributions. In QFT, one has to deal
with an infinite number of many-body amplitudes even though
one may only need to compute a limited number of the
relevant expectation values,
e.g.
$\lan \pi^2 \ran$,  $\lan (\nabla\phi)^2 \ran$, $\lan \phi^2 \ran$,
$\lan \phi^4 \ran$ in $\phi^4$ scalar field theory.
This overcomplicated nature of variational
approach in QFT led us to the following idea.
Instead of going through the complex Fock space representation
of the trial quantum state in computing the expectation values of the relevant
operators, one may parametrize the trial configuration in terms of 
the expectation values themselves, e.g. $\Delta_\pi=\lan \pi^2 \ran$,
$\Delta_{\phi^2} = \lan \phi^2 \ran$ and $\Delta_{\phi^4}=\lan \phi^4 \ran$.
Because of the quantum nature of operators, certain constraints
would be imposed on the expectation values $\Delta$'s by quantum algebra.
Then, the variational problem is cast into a constrained minimization problem
in terms of only the relevant expectation values, e.g.
$$
\ba{l}
\frac 12
(\Delta_{\pi} + \Delta_{\nabla\phi} + m^2 \Delta_{\phi^2}) + 
\frac{\lambda}{4!} \Delta_{\phi^4} \rar \min \\
\text{constraints on }
\Delta_{\pi}, \Delta_{\nabla\phi}, \Delta_{\phi^2}, 
\Delta_{\phi^4}.
\ea
$$

This idea is specifically motivated by the following 
theorem.\newline
{\bf First Symmetric Decomposition 
Theorem:} Symmetrized Fock space ${\cal F}=\{ |\eta\ran \}$ is isomorphic
via
$$
\lan \eta | a_k^\dag a_{k'} | \eta \ran = g(k,k')
$$
to the space of linear integral operators
with kernel $g(k,k')$ that are positive definite hermitian 
and have finite traces.
By this we claim that, if $g(k,k')$  is known
to define a hermitian positive definite linear integral operator with a finite trace,
then there exists a normalized
quantum state in symmetrized Fock space ${\cal F}$ which gives $g(k,k')$
as $\lan a^\dag_k a_{k'}\ran$ and vice versa.

To prove the First Symmetric Decomposition Theorem, we note
that for hermitian operator introduced by kernel $g(k,k')$ there should exist a full
orthonormal
set of eigenfunctions and eigenvalues such that
\be\lab{E12}
g(k,k')=\sum\limits_m \lambda_m g_m(k)g_m^*(k'),
\ee
where $g_m^*(k')$ is a complex conjugation of eigenfunction $g_m(k)$ and
for positive definite operators the eigenvalues $\lambda_m\geq 0$.
It is sufficient to consider a set of states 
\be\lab{E13}
|n\ran = \frac 1{\sqrt {n!}}
\int\!\!\!\int dk_1\dots dk_n f^*_n(k_1,\dots, k_n) a^\dag_{k_1}\dots a^\dag_{k_n} |0\ran,
\ee
where 
\be\lab{E14}
f_n(k_1,\dots, k_n)=\sum\limits_m \sqrt{\lambda_m} g_m(k_1)\dots g_m(k_n).
\ee
Given orthogonality and normalization of the eigenfunctions,
it is easy to check that
$$
\lan n | a^\dag_k a_{k'} | n\ran
=\sum\limits_m \lambda_m g_m(k)g_m^*(k') = g(k,k').
$$
Then, it is always possible to choose a set of amplitudes $\beta_n$ such that
for $| \eta \ran =  \sum\limits_n \beta_n |n\ran$ 
$$
\ba{l}
\lan \eta | a_k^\dag a_{k'} | \eta \ran = g(k,k') \text{ and }
\lan \eta | \eta \ran = 1.
\ea
$$
Note that a finite trace is required provided that
$$
\lan \eta | N | \eta \ran = \text{Tr} [g(k,k')] < \infty.
$$
It is trivial to show that for any $|\eta\ran$ from the
Fock space the expectation value $\lan a_k^\dag a_{k'}\ran$ defines
a positive definite hermitian  linear integral operator with a finite trace.
This completes the proof.

Furthermore, it is  possible to extend this statement to {\bf Second
Symmetric Decomposition Theorem} stating that for
expectation values
on a normalized quantum state from symmetrized Fock space $|\eta\ran$,
\be\lab{E15}
\ba{l}
A_\eta(k,k')=\lan \eta | a_k^\dag a_{k'} + b^\dag_{-k'} b_{-k} |\eta \ran, \\
B_\eta(k,k')=\lan \eta | a_k^\dag b^\dag_{-k'} + a_{k'} b_{-k} |\eta \ran,
\ea
\ee
treated as kernels of hermitian linear integral operators, 
precisely
\be\lab{E16}
1+A_\eta \succeq \sqrt{1+B_\eta^2},
\ee
i.e. $1+A_\eta - \sqrt{1+B_\eta^2}$ is a positive definite 
operator\footnote{note that $B_\eta^2=\lan\eta | B | \eta\ran \lan\eta| B |\eta\ran
\neq \lan\eta | B^2| \eta\ran$ for
$B=a^\dag b^\dag + a b$}.

The significance of these statements is in the claim that any
expectation value of the form given above can be represented in terms
of a state from the Fock space and vice versa. Thus, in dealing with
these expectation values one need not an explicit complex
structure from the Fock space but may work entirely in terms of the 
relevant expectation values constrained by a condition like in Eq.(\ref{E16}).

We will now illustrate application of this principle to the example of
a study of ground state and critical phenomena in a variant of nonlinear
sigma model.
Advantage of this model for our application will be that it is completely formulated
in terms of the expectation values of the quadratic operators in $\phi$ and,
thus, our results about isomorphism between ${\cal F}$ and
($A_\eta$, $B_\eta$) can be straightforwardly applied.
$O(N)$ nonlinear sigma model is introduced as a free field theory constrained to
live on a sphere of radius $R$. It is defined by a free field Hamiltonian
\be\lab{E17}
H = \int dx \frac 12 (\vec \pi^2 +(\nabla \vec \phi)^2 + \mu^2 \vec \phi^2)
\ee
and a constraint for $N$-component vector $\vec \phi(x)$
\be\lab{E18}
|\vec \phi (x)|^2 = R^2.
\ee
In our model we will enforce this constraint softly on average,
so that in the canonical quantization Eqs.(\ref{E17}) and (\ref{E18})  will be given by
\be\lab{E19}
\ba{l}
\lan :H: \ran \sim \int dk \epsilon_k  \sum\limits_i A_\eta^i (k,k) \\
\lan :|\vec \phi^2(x) |:\ran \sim
\int dK dk \frac{e^{ixK}}{\sqrt{\epsilon_k\epsilon_{k+K}}}\sum\limits_i 
[A_\eta^i(k,K+k)+B_\eta^i(k,K+k)]
\ea
\ee
We define the physical subspace as the subset of the Fock space ${\cal F}$
satisfying condition $\lan \eta | :|\phi^2(x)|: | \eta \ran = R^2$.
Consequently, we define the ground state as the state 
from the physical subspace with the lowest energy.

We reformulate an original variational problem for the
ground state in terms of the expectation values
themselves, subject to the quantum algebra constraint (\ref{E16}). In these terms
for the hermitian linear integral operators $A_\eta$ and $B_\eta$, we have
\be\lab{E20}
\left\{
\ba{l}
\text{Tr}[ \epsilon \cdot A_\eta^i]\rar \text{min} \\
\text{Tr}[ \epsilon^{-1} \cdot (A_\eta^i + B_\eta^i)] = R^2 \\
\text{off-diagonal} 
\int dk \frac 1{\sqrt{\epsilon_k\epsilon_{K+k}}} 
\sum\limits_i [A_\eta^i(k,K+k)+B_\eta^i(k,K+k)] = 0,\text{ } K\neq 0.
\ea
\right.
\ee
Here $\epsilon$ is the diagonal matrix with 
$\epsilon_{kk'}=\sqrt{k^2+\mu^2} \delta_{kk'}$,
"$\cdot$" stands for matrix multiplication and Tr stands for integration
over momenta and summation over $i$.
From Eq.(\ref{E16}), we know that for given $B_\eta^i$ linear
operator $A_\eta^i$ 
can be represented by $A_\eta^i = {\cal M}^i + \sqrt{1+(B_\eta^i)^2}-1$,
where ${\cal M}^i \succeq 0$.
Then,
\be\lab{E21}
\ba{l}
\text{Tr}[ \epsilon \cdot A_\eta^i]\rar \min \Leftrightarrow \\
\text{Tr}[ \epsilon \cdot {\cal M}^i] +
\text{Tr}[ \epsilon \cdot (\sqrt{1+(B_\eta^i)^2}-1)]\rar \min.
\ea
\ee
Given ${\cal M}\succeq 0$, diagonal elements
${\cal M}^i_{k,k}\geq 0$ and
Tr$[ \epsilon \cdot { \cal M}^i] \geq 0$ so that to minimize
the expectation value of the Hamiltonian one needs obviously ${\cal M}^i\rar 0$.
Thus, the constraint Eq.(\ref{E16}) is resolved and we get
\be\lab{E22}
\left\{
\ba{l}
\text{Tr}[ \epsilon \cdot (\sqrt{1+(B_\eta^i)^2}-1)]\rar \text{min       (a)} \\
\text{Tr}[ \epsilon^{-1} \cdot (\sqrt{1+(B_\eta^i)^2}-1 + B_\eta^i)] = R^2 \text{  (b)}\\
 \int dk \frac 1{\sqrt{\epsilon_k\epsilon_{K+k}}} 
\sum\limits_i[\sqrt{1+(B_\eta^i)^2}-1+B_\eta^i](k,K+k) = 0,\text{ } K\neq 0 \text{ (c).}
\ea
\right.
\ee
One can observe that for the solution of Eqs.(\ref{E22}a) and (\ref{E22}b)  $B_\eta^i$ 
should be diagonal. For off-diagonal $k\neq k'$, one may consider variation
$\delta B_{kk'}^i$ that preserves Eq.(\ref{E22}b) and show that
\be\lab{E23}
\delta\text{Tr}[\epsilon\cdot(\sqrt{1+(B_\eta^i)^2}-1)]
=(1-\frac{\epsilon_k^2}{\epsilon_{k'}^2})\epsilon_{k'}
\left(\frac {B^i_\eta}{\sqrt{1+(B^i_\eta)^2}}\right)_{k'k} \delta B_{kk'}^i,
\ee
which implies that $B_\eta^i$ with off-diagonal nonzero elements 
cannot be the solution of
the minimization problem (\ref{E22}).
In that case Eq.(\ref{E22}c) is redundant as it is automatically
satisfied by the solution of Eqs.(\ref{E22}a) and (\ref{E22}b).

Having said that, the problem defined by Eqs.(\ref{E22}a) and (\ref{E22}b)
can be solved using the method of Lagrange multipliers. We
introduce a new variable 
\be\lab{E24}
Q_\eta^i(k)=\sqrt{1+(B_\eta^i(k,k))^2}-1+B_\eta^i(k,k),
\ee
and rewrite Eq.(\ref{E22}) in the form
\be\lab{E25}
\ba{l}
\sum\limits_i \int dk \frac{[Q_\eta^i(k)]^2}
{1+Q_\eta^i(k)} \epsilon_k \rar \text{min} \\
\sum\limits_i \int dk Q_\eta^i(k)\epsilon_k^{-1} = R^2,
\ea
\ee
which is rewritten with Lagrange multiplier $\lambda$ as
\be\lab{E26}
\ba{l}
\delta_{Q_\eta^i(k)} 
\sum\limits_i \int dk \left(\frac{[Q_\eta^i(k)]^2}
{1+Q_\eta^i(k)} \epsilon_k - \lambda Q_\eta^i(k)\epsilon_k^{-1} \right) = 0 \Rightarrow \\
Q_\eta^i(k)=\frac{\epsilon_k}{\sqrt{\epsilon_k^2-\lambda(R^2)}} - 1
\ea
\ee
and Eq.(\ref{E25}) is then given by
\be\lab{E27}
N \int \frac{d^{d-1}k}{(2\pi)^{d-1}} \frac 1{2\epsilon_k}
\left[\frac{\epsilon_k}{\sqrt{\epsilon_k^2 - \lambda(R^2)}} - 1 \right] = R^2.
\ee
Eq.(\ref{E27}) is the gap-equation for $\lambda(R^2)$. In 2+1 dimensions,
this integral can be computed exactly to yield
\be\lab{E28}
\frac{N\mu}{4\pi}\left(1-\sqrt{1-\frac{\lambda(R^2)}{\mu^2}}\right) = R^2.
\ee

Thus, we arrive at the following solution of our original problem.
The ground state in our model, defined as the state 
from
the Fock space ${\cal F}$
with
lowest energy $\lan :H: \ran$ and satisfying $\lan :|\phi^2(x)|: \ran = R^2$,
is described for
$R^2\leq R_c^2=\frac{N\mu}{4\pi}$ by the expectation value
$\lan : \phi_i(k)^*\phi_i(k) : \ran = \frac{\epsilon_k}
{\sqrt{\epsilon_k^2 - \lambda}} - 1$, where
$\lambda$ is the solution of Eq.(\ref{E28}).
For $R^2>R_c^2$ the solution of the form (\ref{E26})
can no longer be found because of the "finite capacity" of
$k\neq 0$ modes in the distribution (\ref{E26}). Instead,
the ground state is described as superposition of 
distribution (\ref{E26}) with $\lambda=\mu^2$ and
a singular Bose condensation in $k=0$ mode.
Development of Bose condensation for $R^2>R_c^2$ describes
a phase transition of the second kind in this model.

This discussion is the exact variational solution of the original 
problem thanks 
to the exact constraints on the expectation values $A_\eta$ and $B_\eta$
that we were able to find. 
As one can see, we rendered the explicit
Fock space structure  completely unnecessary. 
The final answer was given in terms of the expectation
values of a given operator on the ground state without explicit reference
to the Fock space.

While the knowledge of an exact image of the Fock space
${\cal F}$ in terms of given expectation values, say
$A_\eta$ and $B_\eta$, is very interesting and, as we have shown,
may be beneficial in certain problems, such detailed
information for the expectation values of more complex
operators, e.g. $\phi^4$, may be difficult to obtain.
In general, if approximate constraints on the expectation 
values of the operators can be established, 
the approach outlined above would provide a variational estimate
for the ground state of the model. 
Among its advantages would be elimination of the necessity to
include the consideration of complicated states
from the Fock space explicitly as well as a possibility to incorporate
renormalizations via the expectation
values of the quantum operators themselves as they enter the final
constrained optimization problem.
In this sense further investigations of this approach
present clear interest.

This work was supported in part by a grant from the U.S. Department of
Energy (DE-FG02-96ER 40947). The National Energy Research Scientific
Computer Center is also acknowledged for the grant of computing time.


\begin{thebibliography}{0}
\bibitem{ritz}
W.~Ritz, J. Reine Phys. Math., {\bf 135}, 1 (1908).
\bibitem{cole} S.~Coleman, Phys. Rev. D {\bf 11}, 2088 (1975);
P.~Stevenson, Phys. Rev. D {\bf 30}, 1712 (1984); Phys. Rev. D {\bf 32},
1389 (1985).
\bibitem{dare} J.~Darewych, M.~Horbatsch, R.~Koniuk, Phys. Rev. D 
{\bf 33}, 3216 (1986); Phys. Rev. Lett. {\bf 54}, 2188 (1985).
\bibitem{dare01} J.~Darewych, A.~Sitenko, I.~Simenog,
A.~Sitnichenko, Phys. Rev. {\bf 147}, 1885 (1995);
L.~Di Leo, J.~Darewych, Can. J. Phys. {\bf 71}, 365 (1993).
\bibitem{tamm} I.~Tamm, J. Phys. USSR {\bf 9}, 449 (1945);
S.~Dancoff, Phys. Rev. {\bf 78}, 382 (1950).

\end{thebibliography}
\end{document}